\documentstyle[12pt,a4wide,times,epsf]{article}

\def\Preprint{\mbox{} 
         \hfill 
 FTUV/97-02 \\ \mbox{}\hfill 
  IFIC/97-02 \\ \mbox{}\hfill January 1997}

\catcode`\@=11

\renewenvironment{thebibliography}[1]
	{\begin{list}{\arabic{enumi}.}
	{\usecounter{enumi}\setlength{\parsep}{0pt}
\setlength{\leftmargin 0.52cm}{\rightmargin 0pt}
	 \setlength{\itemsep}{0pt} \settowidth
	{\labelwidth}{#1.}\sloppy}}{\end{list}}

\def\@cite#1#2{\unskip\nobreak\relax
    \def\@tempa{$\m@th^{\hbox{\the\scriptfont0 #1}}$}%
    \futurelet\@tempc\@citexx}
\def\@citexx{\ifx.\@tempc\let\@tempd=\@citepunct\else
    \ifx,\@tempc\let\@tempd=\@citepunct\else
    \let\@tempd=\@tempa\fi\fi\@tempd}
\def\@citepunct{\@tempc\edef\@sf{\spacefactor=\the\spacefactor\relax}\@tempa
    \@sf\@gobble}

\def\citenum#1{{\def\@cite##1##2{##1}\cite{#1}}}
\def\citea#1{\@cite{#1}{}}

\newcount\@tempcntc
\def\@citex[#1]#2{\if@filesw\immediate\write\@auxout{\string\citation{#2}}\fi
  \@tempcnta\z@\@tempcntb\m@ne\def\@citea{}\@cite{\@for\@citeb:=#2\do
    {\@ifundefined
       {b@\@citeb}{\@citeo\@tempcntb\m@ne\@citea\def\@citea{,}{\bf ?}\@warning
       {Citation `\@citeb' on page \thepage \space undefined}}%
    {\setbox\z@\hbox{\global\@tempcntc0\csname b@\@citeb\endcsname\relax}%
     \ifnum\@tempcntc=\z@ \@citeo\@tempcntb\m@ne
       \@citea\def\@citea{,}\hbox{\csname b@\@citeb\endcsname}%
     \else
      \advance\@tempcntb\@ne
      \ifnum\@tempcntb=\@tempcntc
      \else\advance\@tempcntb\m@ne\@citeo
      \@tempcnta\@tempcntc\@tempcntb\@tempcntc\fi\fi}}\@citeo}{#1}}
\def\@citeo{\ifnum\@tempcnta>\@tempcntb\else\@citea\def\@citea{,}%
  \ifnum\@tempcnta=\@tempcntb\the\@tempcnta\else
   {\advance\@tempcnta\@ne\ifnum\@tempcnta=\@tempcntb \else \def\@citea{--}\fi
    \advance\@tempcnta\m@ne\the\@tempcnta\@citea\the\@tempcntb}\fi\fi}
 
\def\refjl#1#2#3#4#5#6{\bibitem{#1} #2, {\it #3} {#4} (#5) #6.}

\def\etal{{\it et al}}
\def\NP{Nucl. Phys.}
\def\NPPS{Nucl. Phys. B (Proc. Suppl.)}
\def\PL{Phys. Lett.}
\def\PRL{Phys. Rev. Lett.}
\def\PR{Phys. Rev.}
\def\PRep{Phys. Rep.}
\def\ZP{Z. Phys.}

\def\JPG{J. Phys. G: Nucl. Phys.}           

\def\APNY{Ann. Phys., NY}

\def\RPP{Rep. Prog. Phys.}

\newcommand{\eqn}[1]{(\ref{#1})}
\newcommand{\be}{\begin{equation}}
\newcommand{\ee}{\end{equation}}
\newcommand{\no}{\nonumber}
\newcommand{\bel}[1]{\be\label{#1}}
\newcommand{\ba}{\begin{array}{c}}
\newcommand{\bat}{\begin{array}{cc}}
\newcommand{\ea}{\end{array}}
\newcommand{\beqn}{\begin{eqnarray}}
\newcommand{\eeqn}{\end{eqnarray}}

\newcommand{\bi}{\begin{itemize}}
\newcommand{\ei}{\end{itemize}}

\newcommand{\rms}{\rm\scriptsize}

\newcommand{\toLow}{\stackrel{q^2 \ll M_W^2}{\,\longrightarrow\,}}

\newcommand{\ssb}{\stackrel{\mbox{\rm SSB}}{\longrightarrow}}
\newcommand{\cL}{{\cal L}}

\newcommand{\cP}{{\cal P}}

\newcommand{\cH}{{\cal H}}
\newcommand{\cA}{{\cal A}}

\begin{document}
\mbox{}
\noindent\Preprint
\\[3\baselineskip]
\noindent {\bf LEPTON UNIVERSALITY\footnote{Lectures given at the
Carg\`ese'96 School --{\it Masses of Fundamental Particles}--
(Carg\`ese, Corsica, 5--17 August 1996)}}
\\[3\baselineskip] \mbox{}\hspace{2.5cm} 
A. Pich
\\[1\baselineskip] \mbox{}\hspace{2.5cm}
Departament de F\'{\i}sica Te\`orica, IFIC
\\ \mbox{}\hspace{2.5cm} 
CSIC --- Universitat de Val\`encia
\\ \mbox{}\hspace{2.5cm}
Dr. Moliner 50, E--46100 Burjassot, Val\`encia, Spain
\\[1\baselineskip]
\begin{abstract}
\noindent
The Standard Model requires the three known leptonic families to
have identical couplings to the gauge bosons.
The present experimental tests on lepton universality are reviewed,
both for the charged and neutral current sectors.
Our knowledge about the Lorentz structure of the
$l^-\to\nu_l l'^-\bar\nu_{l'}$ transition amplitudes is also
discussed.
\end{abstract}

\vspace{1cm}

\section*{INTRODUCTION}

The Standard Model (SM)
is a gauge theory, based on the group
$SU(3)_C \otimes SU(2)_L \otimes U(1)_Y$,
which describes strong, weak and electromagnetic interactions,
via the exchange of the corresponding spin--1 gauge fields:
8 massless gluons and 1 massless photon for the strong and
electromagnetic interactions, respectively,
and 3 massive bosons, $W^\pm$ and $Z$, for the weak interaction.
The fermionic matter content is given by the known
leptons and quarks, which are organized in a 3--fold
family structure:
\bel{eq:families}
\left[\bat \nu_e & u \\  e^- & d \ea \right] \, , \qquad\quad
\left[\bat \nu_\mu & c \\  \mu^- & s \ea \right] \, , \qquad\quad
\left[\bat \nu_\tau & t \\  \tau^- & b \ea \right] \, , 
\ee
where
(each quark appears in 3 different {\it colours}) 
\bel{eq:structure}
\left[\bat \nu_l & q_u \\  l^- & q_d \ea \right] \,\,\equiv\,\,
\left(\ba \nu_l \\ l^- \ea \right)_{\! L} , \quad
\left(\ba q_u \\ q_d \ea \right)_{\! L} , \quad l^-_R , 
\quad (q_u)_R , \quad
(q_d)_R ,
\ee
plus the corresponding antiparticles.
Thus, the left-handed fields are $SU(2)_L$ doublets, while
their right-handed partners transform as $SU(2)_L$ singlets.
The 3 fermionic families in \eqn{eq:families} appear
to have identical properties (gauge interactions); they only
differ by their mass and their flavour quantum number.

The gauge symmetry is broken by the vacuum,
which triggers the Spontaneous Symmetry Breaking (SSB)
of the electroweak group to the electromagnetic subgroup:
\bel{eq:ssb}
SU(3)_C \otimes SU(2)_L \otimes U(1)_Y \, \ssb\,
SU(3)_C \otimes U(1)_{QED} \, .
\ee
The SSB mechanism generates the masses of the weak gauge bosons,
and gives rise to the appearance of
a physical scalar particle in the model, the so-called {\it Higgs}.
The fermion masses and mixings are also generated through the 
SSB mechanism.  

The SM constitutes one of the most successful achievements
in modern physics. It provides a very elegant theoretical
framework, which is able to describe all known experimental
facts in particle physics.
A detailed description of the SM and its present
phenomenological status can be found in Refs.~\citenum{jaca:94}
and \citenum{sorrento:94}, which discuss the electroweak and strong
sectors, respectively.

In spite of its enormous phenomenological success, the SM leaves too many
unanswered questions to be considered as a complete description of the
fundamental forces.
We do not understand yet why fermions are replicated in three
(and only three)
nearly identical copies? Why the pattern of masses and mixings
is what it is?  Are the masses the only difference among the three
families? What is the origin of the SM flavour structure?
Which dynamics is responsible for the observed CP violation?

The fermionic flavour is the main source of
arbitrary free parameters in the SM: 9 fermion masses,
3 mixing angles and 1 complex phase (assuming the neutrinos to be
massless).
The problem of fermion--mass
generation is deeply related with the mechanism responsible for the SSB.
Thus, the origin of these parameters lies in the most obscure part of
the SM Lagrangian: the scalar sector. 
Clearly, the dynamics of flavour appears to be ``terra incognita''
which deserves a careful investigation.

The flavour structure looks richer in the quark sector, where mixing
phenomena among the different families occur (leptons would also mix
if neutrino masses were non-vanishing).
Since quarks are confined within hadrons, an accurate determination of
their mixing parameters requires first a good understanding of
hadronization effects in flavour--changing transitions.
A rather exhaustive description of our present knowledge on the different
quark  couplings has been given in Ref.~\citenum{comillas:95}.

The leptonic sector is easier to analyze. 
The absence of a direct lepton--gluon vertex 
provides a much cleaner environment to
study the structure of the weak currents and the universality
of their couplings to the gauge bosons.
In the pure leptonic transitions, strong interactions are only
present through small higher--order corrections 
(vacuum polarization, \ldots).
Thus, it is possible to obtain precise theoretical predictions which
can be compared with the available data.
Although hadronization is of course present in
semileptonic decays, such as 
$\tau^-\to\nu_\tau\pi^-$, $\pi^-\to\mu^-\bar\nu_\mu$, \dots,
it only involves
gluonic exchanges between the quarks of a single hadronic current.
Taking appropriate ratios of different semileptonic 
transitions with identical hadronic components, 
the QCD effects cancel to a very good approximation.
Therefore, semileptonic decays also provide
accurate tests of the leptonic couplings. 

\begin{table}
\centering
\caption{Masses and lifetimes of the known leptons
\protect\cite{PDG:96,tau96}.}
\label{tab:masses}
\vspace{0.2cm}
\begin{tabular}{ccc}
\hline
& Mass & Lifetime
\\ \hline
$e$ & $0.51099907\pm 0.00000015$ MeV 
    & $> 4.3\times 10^{23}$ yr
\\
$\mu$ & $105.658389\pm 0.000034$ MeV 
      & $(2.19703\pm0.00004)\times 10^{-6}$ s 
\\ 
$\tau$ & $1777.00{\,}^{+0.30}_{-0.27}$ MeV  
       & $(290.21\pm 1.15)\times 10^{-15}$ s
\\ \hline
$\nu_e$ & $< 10$--15 eV
        & $> 300$ s $\times\, (m_{\nu_e}/$eV) \quad (90\% CL)
\\
$\nu_\mu$ & $< 0.17$ MeV \quad (90\% CL)
          & $> 15.4$ s $\times\, (m_{\nu_\mu}/$eV)  \quad (90\% CL)
\\ 
$\nu_\tau$ & $<18.2$ MeV \quad (95\% CL)
           & Model dependent
\\ \hline
\end{tabular}
\end{table}

The measured masses and lifetimes of the known leptons, shown in
table~\ref{tab:masses}, are very different. The mass spectrum indicates a
hierarchy of the original Yukawa couplings, which increase from one
generation to the other. A similar pattern occurs in the quark sector.
The huge lifetime differences can be simply understood as
a kinematic reflection of the different masses
[see Eq.~\eqn{eq:tau_decay_width}]. 
How precisely we know
that the underlying interactions are actually identical for the 
three lepton generations is the main question we want to address
in the following.

\section*{QED COUPLINGS}

A general description of the electromagnetic coupling of a 
spin--$\frac{1}{2}$ charged lepton to the virtual photon
involves three different form factors:
\bel{eq:em_ff}
T[l\bar l \gamma^*] = e \, \varepsilon_\mu(q) \, \bar l
\left[F_1(q^2)\gamma^\mu
+ i{F_2(q^2)\over 2 m_l} \sigma^{\mu\nu}q_\nu +
{F_3(q^2)\over 2 m_l} \sigma^{\mu\nu}\gamma_5 q_\nu\right] l \ ,
\ee
where $q^\mu$ is the photon momentum.
Owing to the conservation of the electric charge,
$F_1(0)=1$.
At $q^2=0$, the other two form factors reduce to the
lepton 
magnetic dipole moment, 
$\mu_l\equiv (e /2 m_l) \, (g_l/2) = e (1+F_2(0))/2 m_l$,
and electric dipole moment $d_l = e F_3(0)/2 m_l$.

The $F_i(q^2)$ form factors are sensitive quantities to a
possible lepton substructure.
Moreover, $F_3(q^2)$ violates $T$ and $P$ invariance; 
thus, the electric dipole moments, which vanish in the SM, 
constitute a good probe of CP violation.
Owing to their chiral changing structure, the 
magnetic and electric dipole moments 
may provide important insights on the mechanism responsible for
mass generation. In general, one expects \cite{MA:94}
that a fermion of mass
$m_f$ (generated by physics at some scale $M\gg m_f$) will have
induced dipole moments proportional to some power of $m_f/M$.

The measurement of the $e^+e^-\to l^+ l^-$ cross-section has been
used to test the universality of the leptonic QED couplings.
At low energies, where the $Z$ contribution is small, the deviations
from the QED prediction are usually parameterized through\footnote{
A slightly different parameterization is adopted 
for $e^+e^-\to e^+e^-$, to account for
the $t$--channel contribution \protect\cite{MA:90}.}
\be
\sigma(e^+e^-\to l^+ l^-)\, =\, \sigma_{\mbox{\rms QED}} \,
\left( 1 \mp {s\over s-\Lambda_\pm^2}\right)^2.
\ee
The cut-off parameters $\Lambda_\pm$ characterize the validity of QED
and measure the point-like nature of the leptons.
From PEP and PETRA data, one finds \cite{MA:90}:
$\Lambda_+(e)> 435$ GeV, $\Lambda_-(e)> 590$ GeV,
$\Lambda_+(\mu)> 355$ GeV,  $\Lambda_-(\mu)> 265$ GeV,
$\Lambda_+(\tau)> 285$ GeV and 
$\Lambda_-(\tau)> 246$ GeV (95\% CL),
which correspond to upper limits on the lepton charge radii
of about $10^{-3}$ fm.

The most stringent QED test comes of course from the high--precision
measurements of the $e$ and $\mu$ anomalous magnetic moments 
\cite{KI:90,HKS:96,KS:95,BPP:96,KR:96,AL:96}
$a_l\equiv (g_l-2)/2$:
\beqn\label{eq:a_e}
a_e&=&\left\{ \bat
(115 \, 965 \, 214.0\pm 2.8) \times 10^{-11} & (\mbox{\rm Theory})
\\
(115 \, 965 \, 219.3\pm 1.0) \times 10^{-11} & (\mbox{\rm Experiment})
\ea \, , \right.\\ 
a_\mu&=&\left\{ \bat
(1 \, 165 \, 917.1\pm 1.0)   
\times 10^{-9} & (\mbox{\rm Theory})
\\
(1 \, 165 \, 923.0\pm 8.4) \times 10^{-9} & (\mbox{\rm Experiment})
\ea \, . \right.  
\eeqn

Experimentally, very little is known about $a_\tau$ since the spin
precession method used for the lighter leptons cannot be applied
due to the very short lifetime of the $\tau$.
The effect is however visible in the $e^+e^-\to\tau^+\tau^-$
cross-section. The limit $|a_\tau|<0.023$ (95\% CL) has been derived
\cite{SI:83,MA:89}
from PEP and PETRA data. This limit actually probes the corresponding
form factor $F_2(s)$ at $s\sim 35$ GeV.
A more direct bound at $q^2=0$ has been extracted \cite{tau96}
from the decay $Z\to\tau^+\tau^-\gamma$:
\be
|a_\tau| < 0.0104 \qquad (95\%\,\mbox{\rm CL}) \, .
\ee
A slightly better, but more model--dependent, limit has been derived
\cite{EM:93} from the $Z\to\tau^+\tau^-$ decay width:
$-0.004 < a_\tau < 0.006$.

In the SM the overall value of $a_\tau$ is dominated by the second order
QED contribution \cite{SC:48},
$a_\tau \approx \alpha / 2 \pi$.
Including QED corrections up to O($\alpha^3$),
hadronic vacuum polarization contributions
and the corrections due to the weak interactions 
(which are a factor 380
larger than for the muon), the tau anomalous magnetic moment has been
estimated to be \cite{NA:78,SLM:91}
\bel{eq:a_th_tau}
a_\tau\big |_{\mbox{\rms th}} \, = \, (1.1773 \pm 0.0003)
     \times 10^{-3} \, .
\ee

So far, no evidence has been found for any CP--violation signature
in the lepton sector.
The present limits on the leptonic electric dipole moments are
\cite{PDG:96,tau96}:
$$
d_e \, = \,  (-0.3\pm 0.8)\times 10^{-26}\, e\,\mbox{\rm cm},
$$
\be
d_\mu \, = \,  (3.7\pm 3.4)\times 10^{-19}\, e\,\mbox{\rm cm},
\ee
$$
|d_\tau| \, < \,  5.8\times 10^{-17}\, e\,\mbox{\rm cm}.
$$

\section*{CHARGED CURRENT UNIVERSALITY}
\label{sec:cc}

In the SM, the charged--current interactions are governed by an
universal coupling $g$:
\bel{eq:cc_mixing}
\cL_{\mbox{\rms CC}}\, = \, {g\over 2\sqrt{2}}\,\left\{
W^\dagger_\mu\,\left[\sum_{ij}\,
\bar u_i\gamma^\mu(1-\gamma_5) V_{ij} d_j 
\, +\,\sum_l\, \bar\nu_l\gamma^\mu(1-\gamma_5) l
\right]\, + \, \mbox{\rm h.c.}\right\}\, .
\ee
In the original basis of weak eigenstates quarks and leptons have
identical interactions. The diagonalization of the fermion masses
gives rise to the unitary quark mixing matrix $V_{ij}$, which couples
any {\it up--type} quark with all {\it down--type} quarks. 
For massless neutrinos, the analogous leptonic mixing matrix can
be eliminated by a redefinition of the neutrino fields.
The lepton flavour is then conserved in the minimal SM without
right--handed neutrinos.

\subsection*{$\mu^-\to e^-\bar\nu_e\nu_\mu$}

\begin{figure}[bth]
\vfill
\centerline{
\begin{minipage}[t]{.4\linewidth}\centering
{\epsfysize =3.5cm \epsfbox{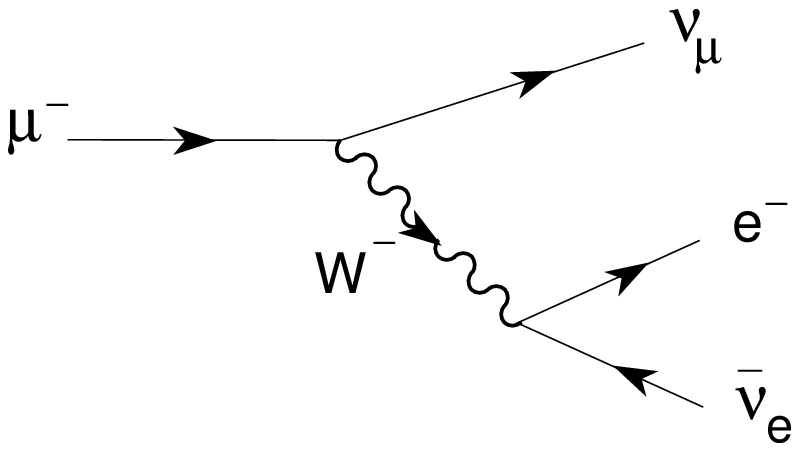}}
\caption{$\mu$-decay diagram.}
\label{fig:mu_decay}
\end{minipage}
\hspace{0.6cm}
\begin{minipage}[t]{.54\linewidth}\centering
{\epsfysize =3.5cm \epsfbox{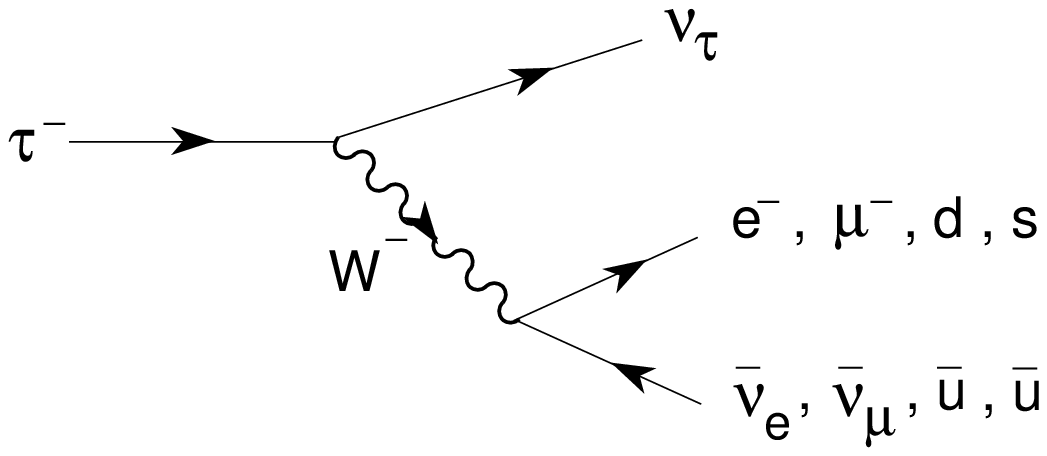}}
\caption{$\tau$-decay diagram.}
\end{minipage}
}
\vfill
\end{figure}

The simplest flavour--changing process is the leptonic
decay of the muon, which proceeds through the $W$--exchange
diagram shown in Fig.~\ref{fig:mu_decay}.
The momentum transfer carried by the intermediate $W$ is very small
compared to $M_W$. Therefore, the vector--boson propagator reduces
to a contact interaction,
\bel{eq:low_energy}
{-g_{\mu\nu} + q_\mu q_\nu/M_W^2 \over q^2-M_W^2}\quad\;
 \toLow\quad\; {g_{\mu\nu}\over M_W^2}\, .
\ee
The decay can then be described through an effective local
4--fermion Hamiltonian,
\bel{eq:mu_v_a}
\cH_{\mbox{\rms eff}}\, = \, {G_F \over\sqrt{2}}
\left[\bar e\gamma^\alpha (1-\gamma_5) \nu_e\right]\,
\left[ \bar\nu_\mu\gamma_\alpha (1-\gamma_5)\mu\right]\, , 
\ee
where
\bel{eq:G_F}
{G_F\over\sqrt{2}} = {g^2\over 8 M_W^2}
\ee
is called the Fermi coupling constant.
$G_F$ 
is fixed by the total decay width,
\bel{eq:mu_lifetime}
{1\over\tau_\mu}\, = \, \Gamma(\mu^-\to e^-\bar\nu_e\nu_\mu)
\, = \, {G_F^2 m_\mu^5\over 192 \pi^3}\,
\left( 1 + \delta_{\mbox{\rms RC}}\right) \, 
f\left(m_e^2/m_\mu^2\right) \, ,
\ee
where
$\, f(x) = 1-8x+8x^3-x^4-12x^2\ln{x}$,
and
\bel{eq:qed_corr}
(1+\delta_{\mbox{\rms RC}})  =  
\left[1+{\alpha(m_\mu)\over 2\pi}\left({25\over 4}-\pi^2\right)\right]\,
\left[ 1 +{3\over 5}{m_\mu^2\over M_W^2} - 2 {m_e^2\over M_W^2}\right]
=  0.9958 \, 
\ee
takes into account the leading higher-order corrections \cite{KS:59,MS:88}.
The measured lifetime \cite{PDG:96},
$\tau_\mu=(2.19703\pm 0.00004)\times 10^{-6}$ s,
implies the value
\bel{eq:gf}
G_F\, = \, (1.16639\pm 0.00002)\times 10^{-5} \:\mbox{\rm GeV}^{-2}
\,\approx\, {1\over (293 \:\mbox{\rm GeV})^2} \, .
\ee
%

\section*{$\tau$ Decay}

The decays of the $\tau$ lepton proceed through the same
$W$--exchange mechanism as the leptonic $\mu$ decay.
The only difference is that several final states
are kinematically allowed:
$\tau^-\to\nu_\tau e^-\bar\nu_e$,
$\tau^-\to\nu_\tau\mu^-\bar\nu_\mu$,
$\tau^-\to\nu_\tau d\bar u$ and $\tau^-\to\nu_\tau s\bar u$.
Owing to the universality of the $W$--couplings, all these
decay modes have equal amplitudes (if final fermion masses and
QCD interactions are neglected), except for an additional
$N_C |V_{ui}|^2$ factor ($i=d,s$) in the semileptonic
channels, where $N_C=3$ is the number of quark colours. 
Making trivial kinematical changes in Eq.~\eqn{eq:mu_lifetime},
one easily gets the lowest--order prediction for the total
$\tau$ decay width:
\bel{eq:tau_decay_width}
{1\over\tau_\tau}\equiv\Gamma(\tau) \approx
\Gamma(\mu) \left({m_\tau\over m_\mu}\right)^5
\left\{ 2 + N_C 
\left( |V_{ud}|^2 + |V_{us}|^2\right)\right\}
\approx {5\over\tau_\mu}\left({m_\tau\over m_\mu}\right)^5,
\ee
where we have used the unitarity relation
$|V_{ud}|^2 + |V_{us}|^2 = 1 - |V_{ub}|^2
\approx 1$. 
From the measured muon lifetime, one has then
$\tau_\tau\approx 3.3\times 10^{-13}$ s, to be compared
with the experimental value \cite{tau96}
$\tau_\tau^{\mbox{\rms exp}} = (2.9021\pm 0.0115)\times 10^{-13}$ s.

%
\begin{table}[thb]
\centering
\caption{Experimental values \protect\cite{tau96}
of some basic $\tau$ decay branching fractions.}
\label{tab:parameters}
\vspace{0.2cm}
\begin{tabular}{lc}
\hline 
$B_e$ & $(17.786\pm 0.072)\% $ \\
$B_\mu$ & $(17.317\pm 0.078)\% $ \\
$R_\tau^B \equiv (1-B_e-B_\mu)/ B_e$ & $3.649\pm 0.019$ \\
Br($\tau^-\to\nu_\tau\pi^-$) & $(11.01\pm 0.11)\% $ \\
Br($\tau^-\to\nu_\tau K^-$) & $(0.692\pm 0.028)\% $ \\
\hline
\end{tabular}
\end{table}
%

The branching ratios into the different decay modes are
predicted to be:
\beqn\label{eq:tau_br}
\mbox{\rm Br}(\tau^-\to\nu_\tau l^-\bar\nu_l) \approx {1\over 5} = 20\%\ , 
\quad\no\\
 R_\tau\equiv {\Gamma(\tau\to\nu_\tau + \mbox{\rm Hadrons})\over 
\Gamma(\tau^-\to\nu_\tau e^-\bar\nu_e)} \approx N_C\, ,
\eeqn
in good agreement with the measured numbers \cite{tau96}, 
given in table~\ref{tab:parameters}. 
Our naive predictions only deviate
from the experimental results by about 20\%. This
is the expected size of the corrections induced by
the strong interactions between the final quarks, 
that we have neglected.
Notice that the measured $\tau$ hadronic width provides strong evidence
for the colour degree of freedom.

The pure leptonic decays 
$\tau^-\to e^-\bar\nu_e\nu_\tau,\mu^-\bar\nu_\mu\nu_\tau$
are theoretically understood at the level of the electroweak
radiative corrections \cite{MS:88}.
The corresponding decay widths are given by Eqs.~\eqn{eq:mu_lifetime}
and \eqn{eq:qed_corr},
making the appropriate changes for the masses of the initial and final
leptons.

Using the value of $G_F$  
measured in $\mu$ decay, Eq.~\eqn{eq:mu_lifetime} 
provides a relation between the $\tau$ lifetime
and the leptonic branching ratios
$B_l\equiv\mbox{\rm Br}(\tau^-\to\nu_\tau l^-\bar\nu_l)$:
\be\label{eq:relation}
B_e =  {B_\mu \over 0.972564\pm 0.000010} = 
{ \tau_{\tau} \over (1.6321 \pm 0.0014) \times 10^{-12}\, \mbox{\rm s} }
\, .
\ee
The errors reflect the present uncertainty of $0.3$ MeV
in the value of $m_\tau$.

\begin{figure}[bth]
\centering
\vspace{0.3cm}
\centerline{\epsfxsize =9cm \epsfbox{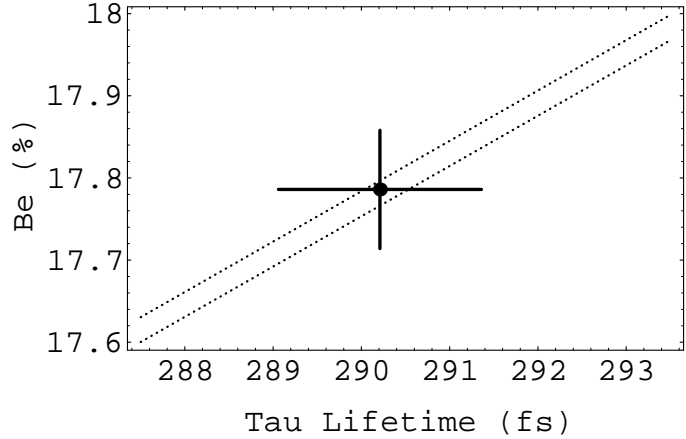}}
\caption{Relation between $B_e$ and $\tau_\tau$. The dotted
band corresponds to the prediction in Eq.~(\protect\ref{eq:relation}).
\label{fig:BeLife}}
\end{figure}

The predicted $B_\mu/B_e$ ratio is in perfect agreement with the measured
value $B_\mu/B_e = 0.974 \pm 0.006$.  As shown in
Fig.~\ref{fig:BeLife}, the relation between $B_e$ and
$\tau_\tau$ is also well satisfied by the present data. Notice, that this
relation is very sensitive to the value of the $\tau$ mass
[$\Gamma(\tau^-\to l^-\bar\nu_l\nu_\tau)\propto m_\tau^5$]. 
The most recent measurements of
$\tau_\tau$, $B_e$ and $m_\tau$ have consistently moved the world averages
in the correct direction, eliminating the previous ($\sim 2\sigma$)
disagreement \cite{PI:92}. 
The experimental precision (0.4\%) is already approaching the
level where a possible non-zero $\nu_\tau$ mass could become relevant; the
present bound \cite{tau96}
$m_{\nu_\tau}< 18.2$ MeV (95\% CL) only guarantees that such 
effect is below 0.08\%.

\section*{Semileptonic Decays}

Semileptonic decays such as $\tau^-\to\nu_\tau P^-$ or 
$P^-\to l^-\bar\nu_l$ [$P=\pi,K$] can be predicted
in a similar way. The effects of the strong interactions
are contained in the so--called decay constants $f_P$,
which parameterize the hadronic matrix element of the
corresponding weak current:
\bel{eq:f_pi_K}\begin{array}{ccc}
\langle\pi^-(p)|\bar d \gamma^\mu\gamma_5 u | 0 \rangle &
\equiv & -i \sqrt{2} f_\pi p^\mu \, , \\
\langle K^-(p)|\bar s \gamma^\mu\gamma_5 u | 0 \rangle &
\equiv & -i \sqrt{2} f_K p^\mu \, . 
\ea\ee

Taking appropriate ratios of different semileptonic decay widths
involving the same meson $P$, the dependence on these decay
constants factors out. Therefore, those ratios can be predicted
rather accurately:
\beqn\label{eq:r_l_P}
R_{e/\mu}& \!\!\!\equiv &\!\!\! {\Gamma(\pi^-\to e^-\bar\nu_e)\over
\Gamma(\pi^-\to \mu^-\bar\nu_\mu)}\, = \,
{m_e^2 (1-m_e^2/m_\pi^2)^2\over m_\mu^2 (1-m_\mu^2/m_\pi^2)^2}
\, (1 + \delta R_{e/\mu})
\, = \, (1.2351\pm 0.0005)\times 10^{-4} ,
\no\\
R_{\tau/\pi} & \!\!\!\equiv &\!\!\!
 {\Gamma(\tau^-\to\nu_\tau\pi^-) \over
 \Gamma(\pi^-\to \mu^-\bar\nu_\mu)} \, = \,
{m_\tau^3\over 2 m_\pi m_\mu^2}
{(1-m_\pi^2/ m_\tau^2)^2\over
 (1-m_\mu^2/ m_\pi^2)^2} 
\left( 1 + \delta R_{\tau/\pi}\right) \, = \,
9774\pm 15 \, ,
\\ 
R_{\tau/K} &\!\!\! \equiv &\!\!\! {\Gamma(\tau^-\to\nu_\tau K^-) \over
 \Gamma(K^-\to \mu^-\bar\nu_\mu)} \, = \,
{m_\tau^3\over 2 m_K m_\mu^2}
{(1-m_K^2/m_\tau^2)^2\over
(1-m_\mu^2/ m_K^2)^2} 
\left( 1 + \delta R_{\tau/K}\right) 
\, = \, 480.4\pm 1.1 \, , \no
\eeqn
where $\delta R_{e/\mu}= - (3.76\pm 0.04)\% $,
$\delta R_{\tau/\pi} = (0.16\pm 0.14)\% $ and
$\delta R_{\tau/K} = (0.90\pm 0.22)\%  $
are the computed \cite{MS:93,DF:94} radiative corrections.
These predictions are in excellent agreement with the
measured ratios \cite{tau96,BR:92,CZ:93}:
$R_{e/\mu} = (1.2310\pm 0.0037)\times 10^{-4}$, 
$R_{\tau/\pi} = 9878\pm 106$ and
$R_{\tau/K} = 465\pm 19$.

\section*{Universality Tests}

All these measurements can be used to test the universality of
the $W$ couplings to the leptonic charged currents.
Allowing the coupling $g$ in Eq.~\eqn{eq:cc_mixing}
to depend on the considered lepton flavour 
(i.e., $g_e$, $g_\mu$, $g_\tau$), 
the ratios $B_\mu/B_e$ and $R_{e/\mu}$ constrain $|g_\mu/g_e|$, while
$B_e/\tau_\tau$ and $R_{\tau/P}$
provide information on $|g_\tau/g_\mu|$.
The present results are shown in tables \ref{tab:univme} and
\ref{tab:univtm}, together with the values obtained from 
the comparison of the $\sigma\cdot B$ partial production
cross-sections for the various $W^-\to l^-\bar\nu_l$ decay
modes at the $p$-$\bar p$ colliders \cite{UA1:89,UA2:92,CDF:92}.

\begin{table}[bth]
\centering
\caption{Present constraints on $|g_\mu/g_e|$.}
\label{tab:univme}
\vspace{0.2cm}
\begin{tabular}{lc}
\hline
& $|g_\mu/g_e|$ \\ \hline
$B_\mu/B_e$ & $1.0005\pm 0.0030$
\\
$R_{\pi\to e/\mu}$ & $1.0017\pm 0.0015$
\\
$\sigma\cdot B_{W\to\mu/e}$ & $1.01\pm 0.04$
\\ \hline
\end{tabular}\vspace{1cm}
%
\caption{Present constraints on $|g_\tau/g_\mu|$.}
\label{tab:univtm}
\vspace{0.2cm}
\begin{tabular}{lc}
\hline
& $|g_\tau/g_\mu|$  \\ \hline
$B_e\tau_\mu/\tau_\tau$ & $1.0001\pm 0.0029$
\\
$R_{\tau/\pi}$ &  $1.005\pm 0.005$
\\
$R_{\tau/K}$ & $0.984\pm 0.020$
\\
$\sigma\cdot B_{W\to\tau/\mu}$ & $0.99\pm 0.05$
\\ \hline
\end{tabular}
\end{table}
%

The present data verify the universality of the leptonic
charged--current couplings to the 0.15\% ($\mu/e$) and 0.30\%
($\tau/\mu$) level. The precision of the most recent
$\tau$--decay measurements is becoming competitive with the 
more accurate $\pi$--decay determination. 
It is important to realize the complementarity of the
different universality tests. 
The pure leptonic decay modes probe
the charged--current couplings of a transverse $W$. In contrast,
the decays $\pi/K\to l\bar\nu$ and $\tau\to\nu_\tau\pi/K$ are only
sensitive to the spin--0 piece of the charged current; thus,
they could unveil the presence of possible scalar--exchange
contributions with Yukawa--like couplings proportional to some
power of the charged--lepton mass.
One can easily imagine new physics scenarios which would modify 
differently the two types of leptonic couplings \cite{MA:94}. 
For instance,
in the usual two Higgs doublet model, charged--scalar exchange
generates a correction to the ratio $B_\mu/B_e$, but 
$R_{\pi\to e/\mu}$ remains unaffected.
Similarly, lepton mixing between the $\nu_\tau$ and an hypothetical
heavy neutrino would not modify the ratios  $B_\mu/B_e$ and
$R_{\pi\to e/\mu}$, but would certainly correct the relation between
$B_l$ and the $\tau$ lifetime.
 
\section*{NEUTRAL CURRENT UNIVERSALITY}
\label{sec:nc}

In the SM, all leptons with equal electric charge have identical
couplings to the $Z$ boson:
\bel{eq:L_nc}
\cL_{\mbox{\rms NC}}^Z \, = \, { g \over 2 \cos{\theta_W}} \,
     Z_\mu \,\sum_l \bar l \gamma^\mu (v_l - a_l \gamma_5) l \, ,
\ee
where 
\be
v_l = T_3^l (1-4|Q_l|\sin^2{\theta_W})\ ,
\qquad\qquad
a_l=T_3^l\, .
\ee
This has been tested at LEP and SLC \cite{LEP:96},
where the {\it effective} vector and axial--vector couplings of the three
charged leptons have been determined.

For unpolarized $e^+$ and $e^-$ beams, the differential 
$e^+e^-\to \gamma,Z\to l^+l^-$
cross-section can be written, at lowest order, as
\bel{eq:dif_cross}
{d\sigma\over d\Omega}\, = \, {\alpha^2\over 8 s} \, 
         \left\{ A \, (1 + \cos^2{\theta}) \, + B\,  \cos{\theta}\,
     - \, h_l \left[ C \, (1 + \cos^2{\theta}) \, +\, D \cos{\theta}
         \right] \right\} ,
\ee
where $h_l$ ($=\pm1$) is the $l^-$ helicity and $\theta$ is 
the scattering angle between $e^-$ and $l^-$.
Here,
\be\label{eq:ABCD}\begin{array}{ccl}
A & = & 1 + 2 v_e v_l \,\mbox{\rm Re}(\chi)
 + \left(v_e^2 + a_e^2\right) \left(v_l^2 + a_l^2\right) |\chi|^2, 
\\
B & = & 4 a_e a_l \,\mbox{\rm Re}(\chi) + 8 v_e a_e v_l a_l  |\chi|^2  , 
\\
C & = & 2 v_e a_l \,\mbox{\rm Re}(\chi) + 2 \left(v_e^2 + a_e^2\right) 
  v_l a_l |\chi|^2 ,
\\
D & = & 4 a_e v_l \,\mbox{\rm Re}(\chi) + 4 v_e a_e \left(v_l^2 +
      a_l^2\right) |\chi|^2  , 
\\ \ea
\ee
and  $\chi$  contains the $Z$  propagator
\bel{eq:Z_propagator}
\chi \, = \, {G_F M_Z^2 \over 2 \sqrt{2} \pi \alpha }
     \,\, {s \over s - M_Z^2 + i s \Gamma_Z  / M_Z } \, . 
\ee

The coefficients $A$, $B$, $C$ and $D$ can be experimentally determined,
by measuring the total cross-section, the forward--backward asymmetry,
the polarization asymmetry and the forward--backward polarization
asymmetry, respectively:
\goodbreak
$$
\sigma(s)  =  {4 \pi \alpha^2 \over 3 s } \, A \, , 
$$
$$
\cA_{\mbox{\rms FB}}(s) \equiv   {N_F - N_B \over N_F + N_B}
     =  {3 \over 8} {B \over A}\,  ,
$$
\be\label{eq:A_pol}
\cA_{\mbox{\rms Pol}}(s)  \equiv
{\sigma^{(h_l =+1)}
- \sigma^{(h_l =-1)} \over \sigma^{(h_l =+1)} + \sigma^{(h_l = -1)}}
\, = \,  - {C \over A} \, ,
\ee
$$
\cA_{\mbox{\rms FB,Pol}}(s)  \equiv  
{N_F^{(h_l =+1)} - 
N_F^{(h_l = -1)} - N_B^{(h_l =+1)} + N_B^{(h_l = -1)} \over
N_F^{(h_l =+1)} + N_F^{(h_l = -1)} + N_B^{(h_l =+1)} + N_B^{(h_l = -1)}}
\, = \, -{3 \over 8} {D \over A}\, . 
$$
Here, $N_F$ and $N_B$ denote the number of $l^-$'s
emerging in the forward and backward hemispheres,
respectively, with respect to the electron direction.

For $s = M_Z^2$,
the real part of the $Z$ propagator vanishes
and the photon exchange terms can be neglected
in comparison with the $Z$--exchange contributions
($\Gamma_Z^2 / M_Z^2 \ll 1$). Eqs.~\eqn{eq:A_pol}
become then,
\beqn
\sigma^{0,l}  \equiv  \sigma(M_Z^2)  = 
 {12 \pi  \over M_Z^2 } \, {\Gamma_e \Gamma_l\over\Gamma_Z^2}\, ,
&& \qquad\;
\cA_{\mbox{\rms FB}}^{0,l}\equiv\cA_{FB}(M_Z^2) = {3 \over 4}
\cP_e \cP_l \, ,
\no\\ \label{eq:A_pol_Z}
\cA_{\mbox{\rms Pol}}^{0,l} \equiv
  \cA_{\mbox{\rms Pol}}(M_Z^2)  = \cP_l \, ,\quad
&& \qquad
\cA_{\mbox{\rms FB,Pol}}^{0,l} \equiv 
\cA_{\mbox{\rms FB,Pol}}(M_Z^2)  =  {3 \over 4} \cP_e  \, ,\quad\quad
\eeqn
where $\Gamma_l$
is the $Z$ partial decay width to the $l^+l^-$ 
final state, and
\bel{eq:P_l}
\cP_l \, \equiv \, { - 2 v_l a_l \over v_l^2 + a_l^2} 
\ee
is the average longitudinal polarization of the lepton $l^-$,
which only depends on the ratio of the vector and axial--vector couplings.
$\cP_l$ is a sensitive function of $\sin^2{\theta_W}$.

The $Z$ partial decay width to the $l^+l^-$ final state,
\bel{eq:Z_l_QED}
\Gamma_l  \equiv\Gamma(Z\to l^+l^-) = 
{G_F M_Z^3\over 6\pi\sqrt{2}} \, (v_l^2 + a_l^2)\, 
\left(1 + {3\alpha\over 4\pi}\right) ,
\ee
determines the sum $(v_l^2 + a_l^2)$, while the ratio $v_l/a_l$
is derived from the asymmetries\footnote{
The asymmetries determine two possible solutions for $|v_l/a_l|$.
This ambiguity can be solved with lower--energy data or
through the measurement of the transverse
spin--spin correlation \cite{BPR:91}
of the two $\tau$'s in $Z\to\tau^+\tau^-$,
which requires \cite{FSanchez} $|v_\tau/a_\tau|<< 1$.}.
The signs of $v_l$ and $a_l$ are fixed by requiring $a_e<0$.

The measurement of the final polarization asymmetries can (only) be done for 
$l=\tau$, because the spin polarization of the $\tau$'s
is reflected in the distorted distribution of their decay products.
Therefore, $\cP_\tau$ and $\cP_e$ can be determined from a
measurement of the spectrum of the final charged particles in the
decay of one $\tau$, or by studying the correlated distributions
between the final products of both $\tau's$ \cite{ABGPR:92}.

With polarized $e^+e^-$ beams, one can also study the left--right
asymmetry between the cross-sections for initial left-- and right--handed
electrons.
At the $Z$ peak, this asymmetry directly measures 
the average initial lepton polarization, $\cP_e$,
without any need for final particle identification:
\bel{eq:A_LR}
\cA_{\mbox{\rms LR}}^0\,\equiv\, \cA_{\mbox{\rms LR}}(M_Z^2)
  \, = \, {\sigma_L(M_Z^2)
- \sigma_R(M_Z^2) \over \sigma_L(M_Z^2) + \sigma_R(M_Z^2)}
\, = \,  - \cP_e \,  .
\ee
%

\begin{table}[tbh]
\centering
\caption{Measured values \protect\cite{LEP:96}
of $\Gamma_l\equiv\Gamma(Z\to l^+l^-)$
and the leptonic forward--backward asymmetries.
The last column shows the combined result 
(for a massless lepton) assuming lepton universality.
\label{tab:LEP_asym}}
\vspace{0.2cm}
\begin{tabular}{ccccc}
\hline
& $e$ & $\mu$ & $\tau$ & $l$ 
\\ \hline
$\Gamma_l$ \, (MeV) & $83.96\pm 0.15$
& $83.79\pm 0.22$ & $83.72\pm 0.26$ & $83.91\pm 0.11$
\\
$\cA_{\mbox{\rms FB}}^{0,l}$ \, (\%) & $1.60\pm 0.24$
& $1.62\pm 0.13$ & $2.01\pm 0.18$ & $1.74\pm 0.10$
\\ \hline \\
\end{tabular}
%
\caption{Measured values \protect\cite{LEP:96}
of the different polarization asymmetries.}
\label{tab:pol_asym}
\vspace{0.2cm}
\begin{tabular}{cccc}
\hline
$\cA_{\mbox{\rms Pol}}^{0,\tau} = \cP_\tau$ &
${4\over 3}\cA^{0,\tau}_{\mbox{\rms FB,Pol}} = \cP_e$ &
$-\cA_{\mbox{\rms LR}}^0 = \cP_e$
& $- \{{4\over 3}\cA_{\mbox{\rms FB}}^{0,l}\}^{1/2} = P_l$
\\ \hline
$-0.1401\pm 0.0067$ & $-0.1382\pm 0.0076$ & $-0.1542\pm 0.0037$
& $-0.1523\pm 0.0044$
\\ \hline
\end{tabular}
\end{table}

\begin{table}[bht]
\centering
\caption{
Effective vector and axial--vector lepton couplings
derived from LEP and SLD data \protect\cite{LEP:96}.
\label{tab:nc_measured}}
\vspace{0.2cm}
\begin{tabular}{ccc}     
\hline
& \multicolumn{2}{c}{Without Lepton Universality}\\ \cline{2-3}
& LEP & LEP + SLD\\ \hline
$v_e$ & $-0.0368\pm 0.0015$ & 
        $-0.03828 \pm 0.00079$
\\
$v_\mu$ & $-0.0372\pm 0.0034$ & 
          $-0.0358 \pm 0.0030$
\\
$v_\tau$ & $-0.0369\pm 0.0016$ & 
           $-0.0367 \pm 0.0016$
\\
$a_e$ & $-0.50130\pm 0.00046$ & 
        $-0.50119 \pm 0.00045$
\\
$a_\mu$ & $-0.50076\pm 0.00069$ & 
          $-0.50086 \pm 0.00068$
\\
$a_\tau$ & $-0.50116\pm 0.00079$ & 
           $-0.50117 \pm 0.00079$
\\ \hline  
$v_\mu/v_e$ & $1.01\pm 0.11$ & 
              $\phantom{-}0.935\pm 0.085$
\\
$v_\tau/v_e$ & $1.001\pm 0.062$ & 
               $\phantom{-}0.959\pm 0.046$
\\
$a_\mu/a_e$ & $0.9989\pm 0.0018$ & 
              $\phantom{-}0.9993\pm 0.0017$
\\
$a_\tau/a_e$ & $0.9997\pm 0.0019$ & 
               $\phantom{-}1.0000\pm 0.0019$
\\ \hline
& \multicolumn{2}{c}{With Lepton Universality}\\ \cline{2-3}
& LEP & LEP + SLD \\ \hline
$v_l$ & $-0.03688\pm 0.00085$ & 
        $-0.03776 \pm 0.00062$
\\
$a_l$ & $-0.50115\pm 0.00034$ & 
        $-0.50108 \pm 0.00034$
\\
$a_\nu=v_\nu$ & $+ 0.5009\pm 0.0010$ & 
                $+0.5009\pm 0.0010$ 
\\ \hline
\end{tabular}
\end{table}

Tables~\ref{tab:LEP_asym} and \ref{tab:pol_asym}
show the present experimental results
for the leptonic $Z$--decay widths and asymmetries.
The data are in excellent agreement with the SM predictions
and confirm the universality of the leptonic neutral couplings\footnote{
A small 0.2\% difference between $\Gamma_\tau$ and $\Gamma_{e,\mu}$
is generated by the $m_\tau$ corrections.}.
There is however a small ($\sim 2\sigma$) discrepancy between the
$\cP_e$ values obtained \cite{LEP:96} from 
$\cA^{0,\tau}_{\mbox{\rms FB,Pol}}$ 
and $\cA_{\mbox{\rms LR}}^0$.
Assuming lepton universality, 
the combined result from all leptonic asymmetries gives
\bel{eq:average_P_l}
\cP_l = - 0.1500\pm 0.0025 \ .
\ee

The measurement of $\cA_{\mbox{\rms Pol}}^{0,\tau}$ and
$\cA^{0,\tau}_{\mbox{\rms FB,Pol}}$ assumes that the $\tau$ decay
proceeds through the SM charged--current interaction.
A more general analysis should take into account the fact that the
$\tau$--decay width depends on the product $\xi\cP_\tau$ 
(see the next section),   
where $\xi$
is the corresponding Michel parameter in leptonic decays, or
the equivalent quantity $\xi_h$  ($=h_{\nu_\tau}$) in the semileptonic 
modes.
A separate measurement of $\xi$ and $\cP_\tau$ has been performed by
ALEPH \cite{ALEPH:94} ($\cP_\tau = -0.139\pm 0.040$)
and L3 \cite{L3:96} ($\cP_\tau = -0.154\pm 0.022$),
using the correlated distribution of the $\tau^+\tau^-$ decays.

\begin{figure}[bht]
\centering
\centerline{\epsfxsize =9cm \epsfbox{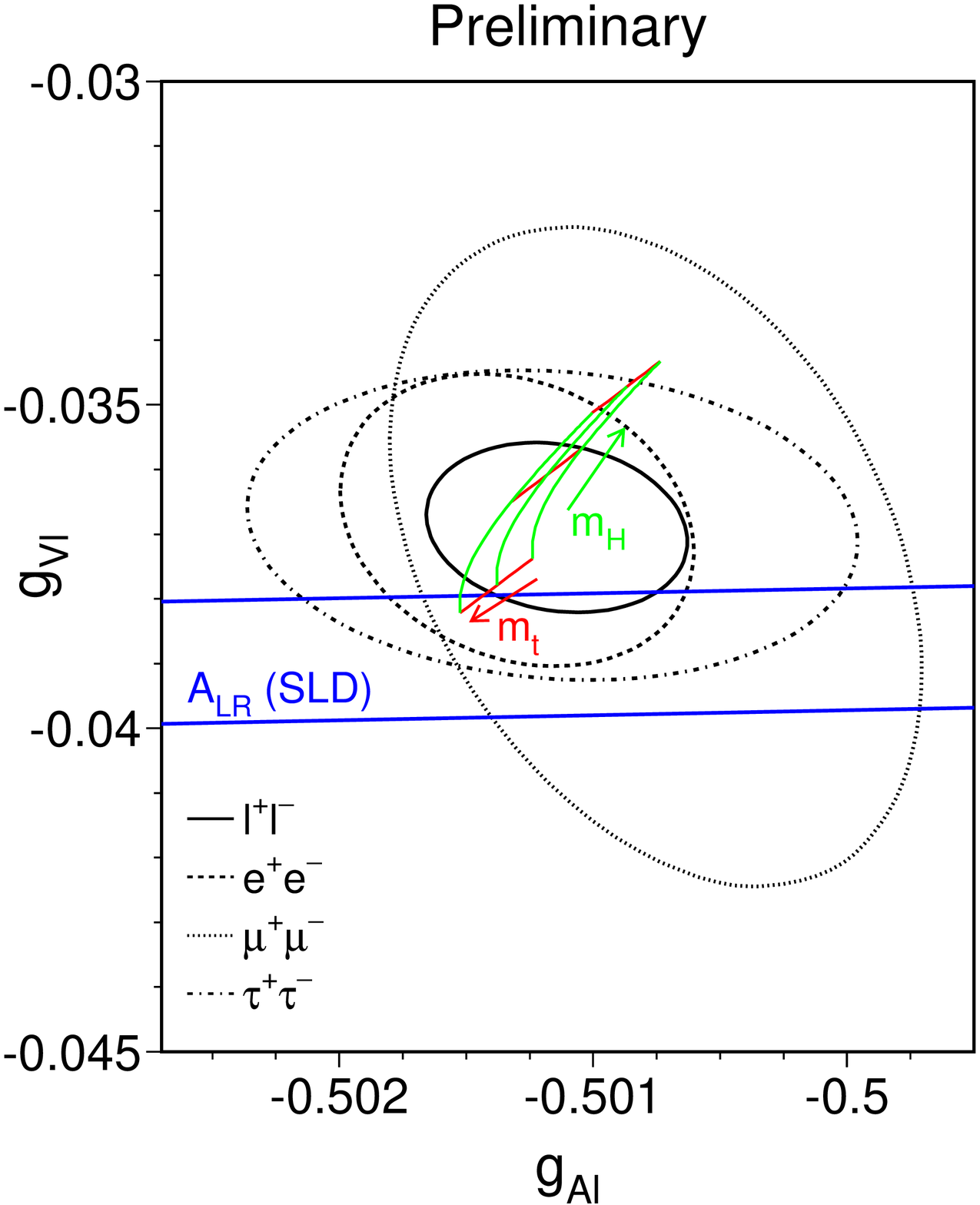}}
\vspace{-0.5cm}
\caption{68\% probability contours in the $a_l$-$v_l$ plane
from LEP measurements \protect\cite{LEP:96}. 
The solid contour assumes lepton universality. 
Also shown is the $1\sigma$ band resulting from the
$\protect\cA_{\mbox{\protect\rms LR}}^0$ measurement at SLD. 
The grid corresponds to the SM prediction.} 
\label{fig:gagv}
\end{figure}

The combined analysis of all leptonic observables from LEP
and SLD ($\cA_{\mbox{\rms LR}}^0$) results in the
effective vector and axial--vector couplings given in
table~\ref{tab:nc_measured} \cite{LEP:96}. 
The corresponding 68\% probability contours in the $a_l$--$v_l$ plane 
are shown in Fig.~\ref{fig:gagv}.
The measured ratios of the $e$, $\mu$ and $\tau$ couplings
provide a test of charged--lepton universality in the neutral--current 
sector.

The neutrino couplings can be determined from the invisible 
$Z$--decay width, by assuming three identical neutrino generations
with left--handed couplings (i.e., $v_\nu=a_\nu$), 
and fixing the sign from neutrino scattering 
data \cite{CHARMII:94}.
The resulting experimental value \cite{LEP:96},
given in table~\ref{tab:nc_measured},
is in perfect agreement with the SM.
Alternatively, one can use the SM prediction for 
$\Gamma_{\mbox{\rms inv}}/\Gamma_l$
to get a determination of the number of (light) neutrino flavours
\cite{LEP:96}:
\be
N_\nu = 2.989\pm 0.012 \, .
\ee
The universality of the neutrino couplings has been tested
with $\nu_\mu e$ scattering data, which fixes \cite{CHARMII:94b}
the $\nu_\mu$ coupling to the $Z$: \ 
$v_{\nu_\mu} =  a_{\nu_\mu} = 0.502\pm 0.017$.

The measured leptonic asymmetries can be used to obtain the
effective electroweak mixing angle in the charged--lepton sector: 
\cite{LEP:96}
\bel{eq:bar_s_W_l}
\sin^2{\theta^{\mbox{\rms lept}}_{\mbox{\rms eff}}} \equiv
{1\over 4}   \left( 1 - {v_l\over a_l}\right)  = 0.23114\pm 0.00031 \, .
\ee
Including also the hadronic asymmetries, one gets \cite{LEP:96}
$\sin^2{\theta^{\mbox{\rms lept}}_{\mbox{\rms eff}}} =
0.23165\pm 0.00024$ 
with a $\chi^2/\mbox{\rm d.o.f.} = 12.8/6$.

\section*{LORENTZ STRUCTURE OF THE CHARGED CURRENTS}
\label{sec:lorentz}

Let us consider the 
decay $l^-\to\nu_l l'^-\bar\nu_{l'}$, 
where the lepton pair ($l$, $l^\prime $)
may be ($\mu$, $e$), ($\tau$, $e$), or ($\tau$, $\mu$). 
The most general, local, derivative--free, lepton--number conserving, 
four--lepton interaction Hamiltonian, 
consistent with locality and Lorentz invariance
\cite{MI:50,BM:57,SCH:83,FGJ:86,FG:93,PS:95},
\be
{\cal H} = 4 \frac{G_{l'l}}{\sqrt{2}}
\sum_{n,\epsilon,\omega}          
g^n_{\epsilon\omega}   
\left[ \overline{l'_\epsilon} 
\Gamma^n {(\nu_{l'})}_\sigma \right]\, 
\left[ \overline{({\nu_l})_\lambda} \Gamma_n 
	l_\omega \right]\ ,
\label{eq:hamiltonian}
\ee
contains ten complex coupling constants or, since a common phase is
arbitrary, nineteen independent real parameters
which could be different for each leptonic decay.
The subindices
$\epsilon , \omega , \sigma, \lambda$ label the chiralities (left--handed,
right--handed)  of the  corresponding  fermions, and $n$ the
type of interaction: 
scalar ($I$), vector ($\gamma^\mu$), tensor 
($\sigma^{\mu\nu}/\sqrt{2}$).
For given $n, \epsilon ,
\omega $, the neutrino chiralities $\sigma $ and $\lambda$
are uniquely determined.

Taking out a common factor $G_{l'l}$, which is determined by the total
decay rate, the coupling constants $g^n_{\epsilon\omega}$
are normalized to \cite{FGJ:86}
\beqn\label{eq:normalization}
1 &\!\!\! = &\!\!\!
{1\over 4} \,\left( |g^S_{RR}|^2 + |g^S_{RL}|^2
    + |g^S_{LR}|^2 + |g^S_{LL}|^2 \right)
    +  3 \,\left( |g^T_{RL}|^2 + |g^T_{LR}|^2 \right) 
\no \\ & &\!\!\! \mbox{}
+ \left(
   |g^V_{RR}|^2 + |g^V_{RL}|^2 + |g^V_{LR}|^2 + |g^V_{LL}|^2 \right)
\, .
\eeqn
In the SM, $g^V_{LL}  = 1$  and all other
$g^n_{\epsilon\omega} = 0 $.

For an initial lepton polarization ${\cal P}_l$,
the final charged--lepton distribution in the decaying--lepton 
rest frame
is usually parameterized \cite{BM:57} in the form  
\be\label{eq:spectrum}
{d^2\Gamma \over dx\, d\cos\theta} =
{m_l\omega^4 \over 2\pi^3} G_{l'l}^2 \sqrt{x^2-x_0^2}
\left\{ F(x) - 
{\xi\over 3}\, {\cal P}_l\,\sqrt{x^2-x_0^2}
\,\cos{\theta}\, A(x)\right\} ,
\ee
where $\theta$ is the angle between the $l^-$ spin and the
final charged--lepton momentum,
$\, \omega \equiv (m_l^2 + m_{l'}^2)/2 m_l \, $
is the maximum $l'^-$ energy for massless neutrinos, $x \equiv E_{l'^-} /
\omega$ is the reduced energy, $x_0\equiv m_{l'}/\omega$
and
\beqn\label{eq:Fx_Ax_def}
F(x)  &\!\!\! = &\!\!\! 
  x (1 - x) + {2\over 9} \rho 
 \left(4 x^2 - 3 x - x_0^2 \right) 
+  \eta\, x_0 (1-x)
\, , \no\\
A(x) &\!\!\! = &\!\!\! 
 1 - x   + {2\over 3}  \delta \left( 4 x - 4 + \sqrt{1-x_0^2}  
\right)  \, .
\eeqn

For unpolarized $l's$, the distribution is characterized by
the so-called Michel \cite{MI:50} parameter $\rho$
and the low--energy parameter $\eta$. Two more parameters, $\xi$
and $\delta$, can be determined when the initial lepton polarization is known.
If the polarization of the final charged lepton is also measured,
5 additional independent parameters \cite{PDG:96}  
($\xi'$, $\xi''$, $\eta''$, $\alpha'$, $\beta'$)
appear. 

For massless neutrinos, the total decay rate is given by \cite{PS:95}
\be\label{eq:gamma}
\Gamma\, = \, {m_l^5 \widehat{G}_{l'l}^2\over 192 \pi^3}\,
f\!\left({m_{l'}^2\over m_l^2}\right)\, 
(1 + \delta_{\mbox{\rms RC}})
\, ,
\ee
where
\be
\widehat{G}_{l'l}  \equiv  G_{l'l} \,
\sqrt{1 + 4\,\eta\, {m_{l'}\over m_l}\,
{g\!\left( m_{l'}^2/ m_l^2 \right)\over  
f\!\left( m_{l'}^2/ m_l^2 \right)}}
\, ,
\ee
and $g(z) = 1 + 9 z - 9 z^2 - z^3 + 6 z (1+z) \ln{z}$.
Thus, $\widehat{G}_{e\mu}$ corresponds to the Fermi coupling 
$G_F$, measured in $\mu$ decay.
The $B_\mu/B_e$ and $B_e\tau_\mu/\tau_\tau$
universality tests, discussed in the previous section,
actually prove the ratios
$|\widehat{G}_{\mu\tau}/\widehat{G}_{e\tau}|$
and $|\widehat{G}_{e\tau}/\widehat{G}_{e\mu}|$, respectively.
An important point, emphatically stressed by
Fetscher and Gerber \cite{FG:93}, concerns the extraction
of $G_{e \mu}$, whose uncertainty is dominated
by the uncertainty in $\eta_{\mu\to e}$. 

In terms of the $g_{\epsilon\omega}^n$
couplings, the shape parameters in Eqs.~\eqn{eq:spectrum}
and \eqn{eq:Fx_Ax_def}
are:
\beqn\label{eq:michel}
\rho &\!\!\! = &\!\!\! 
{3\over 4} (\beta^+ + \beta^-) + (\gamma^+ + \gamma^-) \, ,
\no\\
\xi &\!\!\! = &\!\!\! 3 (\alpha^- - \alpha^+) + (\beta^- - \beta^+)
  + {7\over 3} (\gamma^+ - \gamma^-) \, ,
\no\\
\xi\delta &\!\!\! = &\!\!\! 
{3\over 4} (\beta^- - \beta^+) + (\gamma^+ - \gamma^-) \, ,
\\
\eta &\!\!\! = &\!\!\! 
\frac{1}{2} \mbox{\rm Re}\left[
g^V_{LL} g^{S\ast}_{RR} + g^V_{RR}  g^{S\ast}_{LL}
+ g^V_{LR} \left(g^{S\ast}_{RL} + 6 g^{T\ast}_{RL}\right)
+ g^V_{RL} \left(g^{S\ast}_{LR} + 6 g^{T\ast}_{LR}\right) 
\right] , \no
\eeqn
where \cite{Rouge}
%
%
\beqn\label{eq:abg_def}
\lefteqn{\alpha^+ \equiv  
{|g^V_{RL}|}^2 + {1\over 16} {|g^S_{RL} + 6 g^T_{RL}|}^2
\, , } && \qquad\qquad\qquad\qquad\qquad\qquad\qquad\qquad\qquad
\alpha^- \equiv  
{|g^V_{LR}|}^2 + {1\over 16} {|g^S_{LR} + 6 g^T_{LR}|}^2
\, , 
\no\\
\lefteqn{\beta^+ \equiv 
 {|g^V_{RR}|}^2 + {1\over 4} {|g^S_{RR}|}^2
\, , } && \qquad\qquad\qquad\qquad\qquad\qquad\qquad\qquad\qquad
\beta^- \equiv 
 {|g^V_{LL}|}^2 + {1\over 4} {|g^S_{LL}|}^2
\, ,
\\
\lefteqn{\gamma^+ \equiv 
 {3\over 16} {|g^S_{RL} - 2 g^T_{RL}|}^2
\, , } && \qquad\qquad\qquad\qquad\qquad\qquad\qquad\qquad\qquad
\gamma^- \equiv 
 {3\over 16} {|g^S_{LR} - 2 g^T_{LR}|}^2
\, ,
\no
\eeqn
are positive--definite combinations of decay constants, corresponding to 
a final right-- ($\alpha^+$, $\beta^+$, $\gamma^+$) or
left-- ($\alpha^-$, $\beta^-$, $\gamma^-$)
handed lepton. 
%
In the SM, $\rho = \delta = 3/4$, 
$\eta = \eta'' = \alpha' = \beta' = 0 $ and 
$\xi = \xi' = \xi'' = 1 $.

The normalization constraint \eqn{eq:normalization} is equivalent to
$\alpha^+ + \alpha^- + \beta^+ + \beta^- + \gamma^+ + \gamma^- = 1$.
It is convenient to introduce \cite{FGJ:86} the probabilities
$Q_{\epsilon\omega}$ for the
decay of an $\omega$--handed $l^-$
into an $\epsilon$--handed 
daughter lepton,
\begin{eqnarray}\label{eq:Q_LL}
Q_{LL} &\!\!\! = &\!\!\!
{1 \over 4} |g^S_{LL}|^2 \! +  |g^V_{LL}|^2 
\phantom{+ 3 |g^T_{LR}|^2} 
 = {1 \over 4}\left(
-3 +{16\over 3}\rho -{1\over 3}\xi +{16\over 9}\xi\delta +\xi'+\xi''
\right)\! , \quad\;\;\no\\ 
Q_{RR} &\!\!\! = &\!\!\! 
{1 \over 4} |g^S_{RR}|^2 \! + \! |g^V_{RR}|^2 
\phantom{+ 3 |g^T_{LR}|^2} 
 =  {1 \over 4}\left(
-3 +{16\over 3}\rho +{1\over 3}\xi -{16\over 9}\xi\delta -\xi'+\xi''
\right)\!  , \no\\ 
Q_{LR} &\!\!\! = &\!\!\! 
{1 \over 4} |g^S_{LR}|^2 \! + \!  |g^V_{LR}|^2
            \! + \!   3 |g^T_{LR}|^2  
 = {1 \over 4}\left(
5 -{16\over 3}\rho +{1\over 3}\xi -{16\over 9}\xi\delta +\xi'-\xi''
\right)\! , \\ 
Q_{RL} &\!\!\! = &\!\!\! 
{1 \over 4} |g^S_{RL}|^2  \! + \!  |g^V_{RL}|^2
            \! + \!  3 |g^T_{RL}|^2  
= {1 \over 4}\left(
5 -{16\over 3}\rho -{1\over 3}\xi +{16\over 9}\xi\delta -\xi'-\xi''
\right)\! . \no
\end{eqnarray}
Upper bounds on any of these (positive--semidefinite) probabilities 
translate into corresponding limits for all couplings with the 
given chiralities.

For $\mu$ decay, where precise measurements of the polarizations of
both $\mu$ and $e$ have been performed, there exist \cite{FGJ:86}
upper bounds on $Q_{RR}$, $Q_{LR}$ and $Q_{RL}$, and a lower bound
on $Q_{LL}$. They imply corresponding upper bounds on the 8
couplings $|g^n_{RR}|$, $|g^n_{LR}|$ and $|g^n_{RL}|$.
The measurements of the $\mu^-$ and the $e^-$ do not allow to
determine $|g^S_{LL}|$ and $|g^V_{LL}|$ separately \cite{FGJ:86,JA:66}.
Nevertheless, since the helicity of the $\nu_\mu$ in pion decay is
experimentally known \cite{FE:84}    
to be $-1$, a lower limit on $|g^V_{LL}|$ is
obtained \cite{FGJ:86} from the inverse muon decay
$\nu_\mu e^-\to\mu^-\nu_e$.
The present (90\% CL) bounds \cite{PDG:96}
on the $\mu$--decay couplings
are shown in Fig.~\ref{fig:mu_couplings}. 
These limits show nicely 
that the bulk of the $\mu$--decay transition amplitude is indeed of
the predicted V$-$A type.

\begin{figure}[bth]
\vfill
\centerline{
\begin{minipage}[t]{.47\linewidth}\centering
\centerline{\epsfxsize =8cm \epsfbox{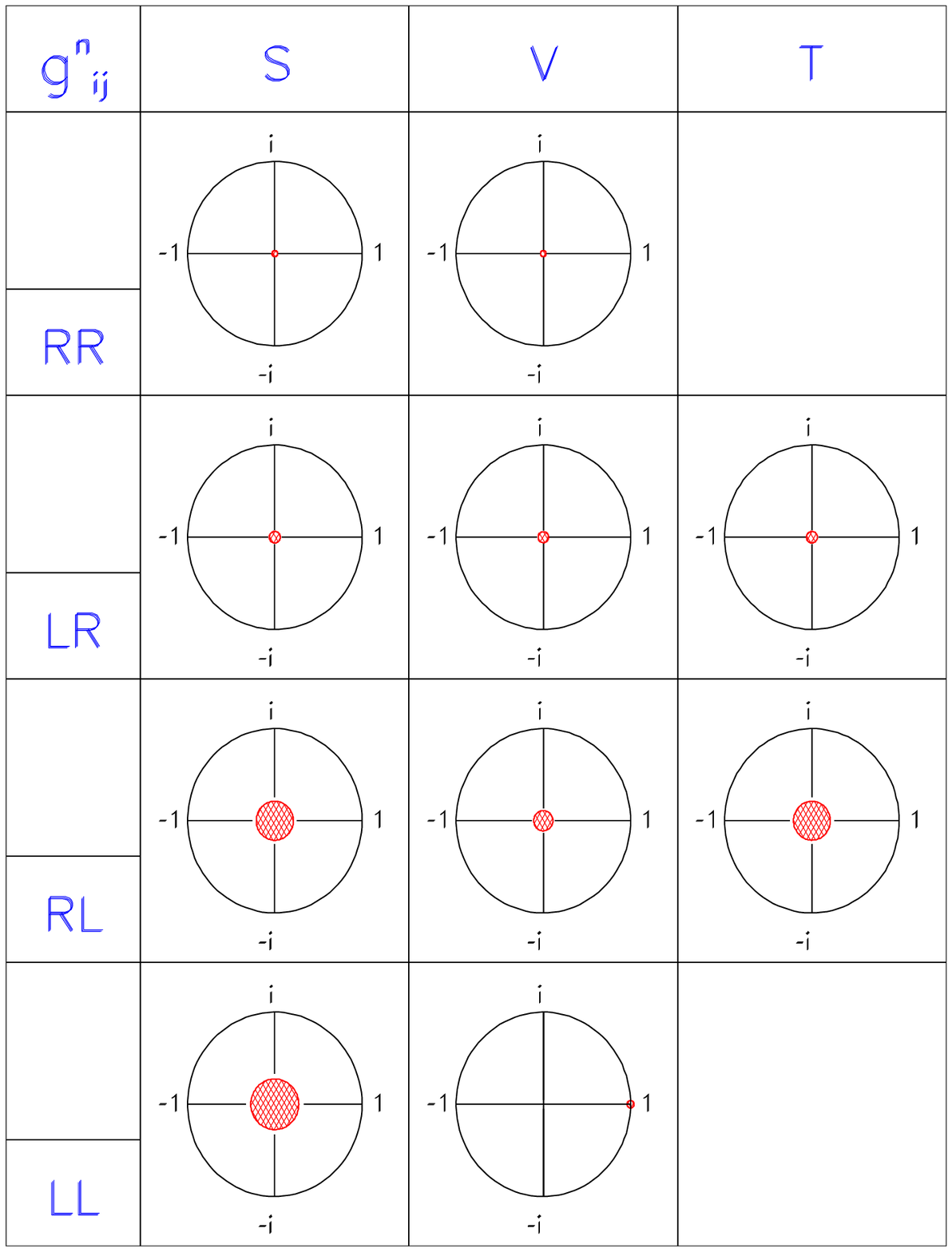}}
\vspace{-1.5cm}
\caption{90\% CL experimental limits  \protect\cite{PDG:96} 
for the normalized $\mu$--decay couplings
$g'^n_{\epsilon\omega }\equiv g^n_{\epsilon\omega }/ N^n$,
where
$N^n \equiv \protect\mbox{\rm max}(|g^n_{\epsilon\omega }|) =2$,
1, $1/\protect\sqrt{3} $ for $n =$ S, V, T.
(Taken from Ref.~\protect\citenum{LR:95}).}
\label{fig:mu_couplings}
\end{minipage}
\hspace{0.936cm}
\begin{minipage}[t]{.47\linewidth}\centering
\centerline{\epsfxsize =8cm \epsfbox{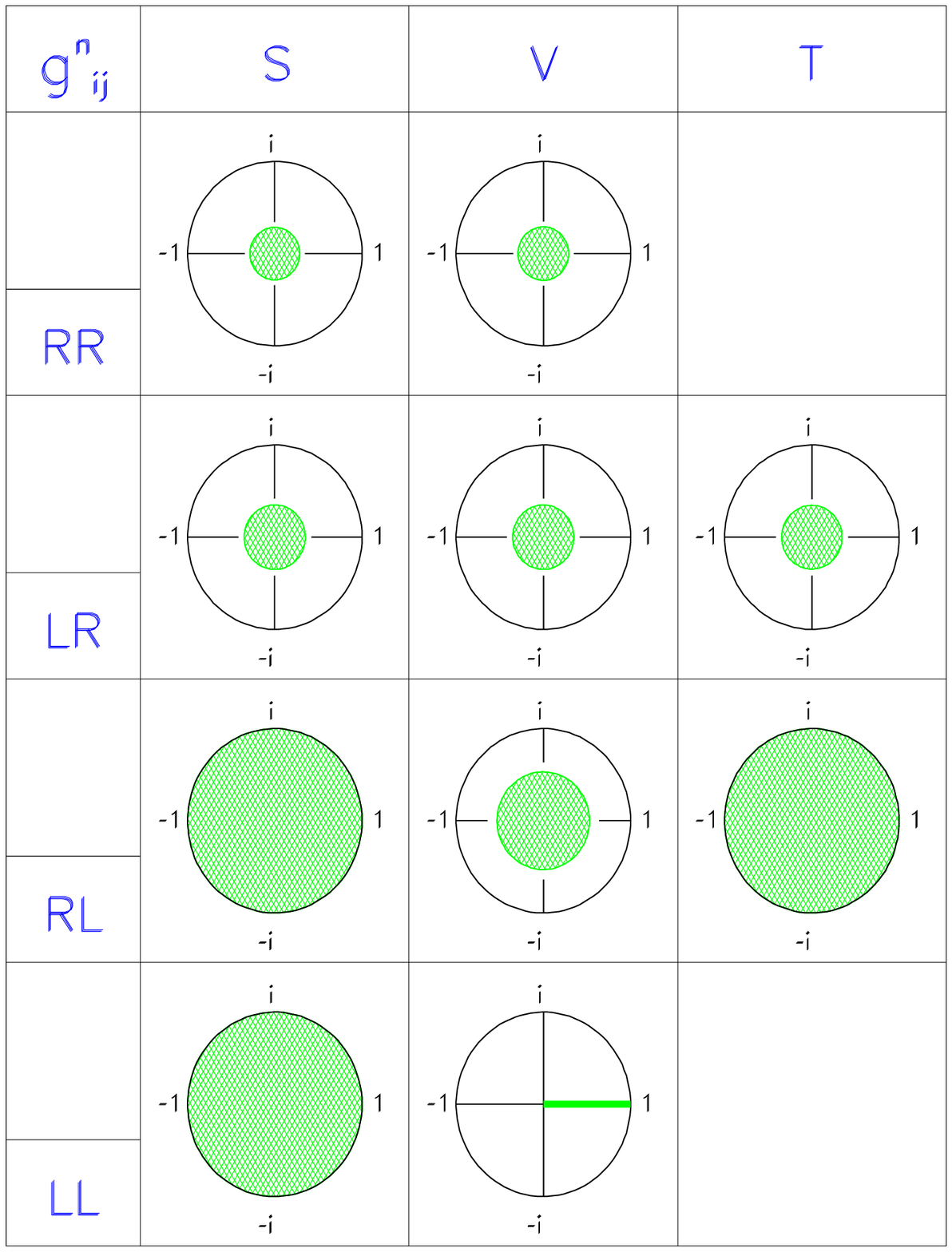}}
\vspace{-1.5cm}
\caption{90\% CL experimental limits
for the normalized $\tau$--decay couplings
$g'^n_{\epsilon\omega }\equiv g^n_{\epsilon\omega }/ N^n$,
assuming $e/\mu$ universality.
\label{fig:tau_couplings}}
\end{minipage}
}
\vfill
\end{figure}

The experimental analysis of the $\tau$--decay parameters is 
necessarily
different from the one applied to the muon, because of the much
shorter $\tau$ lifetime.
The measurement of the $\tau$ polarization and the parameters
$\xi$ and $\delta$ 
is possible due to the fact that the spins
of the $\tau^+\tau^-$ pair produced in $e^+e^-$ annihilation 
are strongly correlated
\cite{BPR:91,ABGPR:92,TS:71,PS:77,NE:91,GN:91,FE:90,DDDR:93}.
Another possibility is to use
the beam polarization, as done by SLD.
However,
the polarization of the charged lepton emitted in the $\tau$ decay
has never been measured. In principle, this could be done
for the decay $\tau^-\to\mu^-\bar\nu_\mu\nu_\tau$ by stopping the
muons and detecting their decay products \cite{FE:90}.
An alternative method would be \cite{SV:96} to use 
the radiative decays
$\tau\to l^-\bar\nu_l\nu_\tau\gamma$ ($l=e,\mu$), 
since the
distribution of the photons emitted by the daughter lepton is
sensitive to the lepton spin. 
The measurement of the inverse decay $\nu_\tau l^-\to\tau^-\nu_l$
looks far out of reach.

The present experimental status \cite{tau96}
on the $\tau$--decay Michel parameters
is shown in table~\ref{tab:tau_michel}.
For comparison, the values measured in $\mu$ decay \cite{PDG:96}
are also given.
The improved accuracy of the most recent experimental analyses
has brought an enhanced sensitivity to the different shape parameters,
allowing the first measurements \cite{tau96} of $\eta_{\tau\to\mu}$, 
$\xi_{\tau\to e}$, $\xi_{\tau\to\mu}$, $(\xi\delta)_{\tau\to e}$ and 
$(\xi\delta)_{\tau\to\mu}$
without any $e/\mu$ universality assumption.

\begin{table}[bth]
\centering
\caption{World average
\protect\cite{PDG:96,tau96}
Michel parameters. 
The last column ($\tau\to l$) assumes identical couplings
for $l=e,\mu$.
$\xi_{\mu\to e}$ refers to the product $\xi_{\mu\to e}\cP_\mu$,
where $\cP_\mu\approx 1$ is the longitudinal polarization
of the $\mu$ from $\pi$ decay.}
\label{tab:tau_michel}
\vspace{0.2cm}
\begin{tabular}{lcccc}
\hline
& $\mu\to e$ & $\tau\to\mu$ & $\tau\to e$ & $\tau\to l$ 
\\ \hline
$\rho$ & $0.7518\pm 0.0026$ & $0.733\pm 0.031$ & $0.734\pm 0.016$ & 
$0.741\pm 0.014$ 
\\
$\eta$ & $-0.007\pm 0.013\phantom{-}$ & $-0.04\pm 0.20\phantom{-}$ & --- & 
$0.047\pm 0.076$
\\
$\xi$ & $1.0027\pm 0.0085$ & $1.19\pm 0.18$ & $1.09\pm 0.16$ & 
$1.04\pm 0.09$ 
\\
$\xi\delta$ & $0.7506\pm 0.0074$ & $0.73\pm 0.11$ & $ 0.80\pm 0.18$ & 
$ 0.73\pm 0.07$ 
\\ \hline \\
\end{tabular}
\end{table}

The determination of the $\tau$ polarization parameters 
allows us to bound the total probability for the decay of
a right--handed $\tau$ \cite{FE:90},
\be\label{eq:Q_R}
Q_{\tau_R} \equiv 
Q_{RR} + Q_{LR}
= \frac{1}{2}\, \left[ 1 + \frac{\xi}{3} - \frac{16}{9} 
(\xi\delta)\right]
\; .
\ee
One finds (ignoring possible correlations among the measurements):
\begin{eqnarray}
Q_{\tau_R}^{\tau\to\mu} &\!\!\! =&\!\!\! \phantom{-}0.05\pm 0.10 \; 
< \, 0.20 \quad (90\%\;\mbox{\rm CL})\, , \no\\
Q_{\tau_R}^{\tau\to e} &\!\!\! =&\!\!\! -0.03\pm 0.16 \; 
< \, 0.25 \quad (90\%\;\mbox{\rm CL})\, , \\
Q_{\tau_R}^{\tau\to l} &\!\!\! =&\!\!\! \phantom{-}0.02\pm 0.06 \;
< \, 0.12 \quad (90\%\;\mbox{\rm CL})\, , \no
\end{eqnarray}
where the last value refers to the $\tau$ decay into either $l=e$ or $\mu$,
assuming identical $e$/$\mu$ couplings.
Since these probabilities are positive semidefinite quantities, they imply
corresponding limits on all
$|g^n_{RR}|$ and $|g^n_{LR}|$ couplings. 

A measurement of the final lepton polarization could be even more efficient,
since the total probability for the decay into a right--handed lepton
depends on a single Michel parameter:
\bel{eq:Qxi}
Q_{l'_R} \equiv Q_{RR} +  Q_{RL}
= {1\over 2} ( 1 -\xi') \, .
\ee
Thus, a single polarization measurement could bound the five RR and RL
complex couplings.

Another useful positive--definite quantity is \cite{LR:95}
\bel{eq:rxd}
\rho - \xi\delta = {3\over 2} \beta^+ + 2 \gamma^- \, ,
\ee
which provides direct bounds on $|g^V_{RR}|$ and $|g^S_{RR}|$.
A rather weak upper limit on $\gamma^+$ is obtained from the
parameter $\rho$. More stringent is the bound on $\alpha^+$ obtained from
$(1-\rho)$, which is also positive--definite; it implies a corresponding
limit on $|g^V_{RL}|$.

Table~\ref{table:g_tau_bounds} gives the resulting (90\% CL) bounds on the 
$\tau$--decay couplings.
The relevance of these limits can be better appreciated in 
Fig.~\ref{fig:tau_couplings}, 
where $e$/$\mu$ universality has been assumed.

\begin{table}[bth]
\centering
\caption{90\% CL limits
for the $g^n_{\epsilon\omega}$ couplings.}
\label{table:g_tau_bounds}
\vspace{0.2cm}
\begin{tabular}{lllll}
\hline
& \hfil $\mu\to e$\hfil &
\hfil $\tau\to\mu$\hfil &\hfil $\tau\to e$ \hfil & 
\hfil $\tau\to l$ \hfil
\\ \hline
$|g^S_{RR}|$  & $< 0.066$ & $< 0.71$ & $< 0.83$ & $< 0.57$
\\
$|g^S_{LR}|$  & $< 0.125$ & $< 0.90$ & $< 1.00$ & $< 0.70$
\\
$|g^S_{RL}|$  & $< 0.424$ & $\leq 2$ & $\leq 2$ & $\leq 2$
\\
$|g^S_{LL}|$  & $< 0.55$  & $\leq 2$ & $\leq 2$ & $\leq 2$
\\ \hline
$|g^V_{RR}|$  & $< 0.033$ & $< 0.36$ & $< 0.42$ & $< 0.29$
\\
$|g^V_{LR}|$  & $< 0.060$ & $< 0.45$ & $< 0.50$ & $< 0.35$
\\
$|g^V_{RL}|$  & $< 0.110$ & $< 0.56$ & $< 0.54$ & $< 0.53$
\\
$|g^V_{LL}|$  & $> 0.96$  & $\leq 1$ & $\leq 1$ & $\leq 1$
\\ \hline
$|g^T_{LR}|$  & $< 0.036$ & $< 0.26$ & $< 0.29$ & $< 0.20$
\\
$|g^T_{RL}|$  & $< 0.122$ & $\leq 1/\sqrt{3}$ & $\leq 1/\sqrt{3}$
              & $\leq 1/\sqrt{3}$
\\ \hline
\end{tabular}
\end{table}

If lepton universality is assumed, 
the leptonic decay ratios $B_\mu/B_e$  and $B_e\tau_\mu/\tau_\tau$
provide limits on the low--energy parameter $\eta$.  
The best sensitivity \cite{ST:94} comes from
$\widehat{G}_{\mu\tau}$,
where the term proportional to $\eta$ is not suppressed by
the small $m_e/m_l$ factor. The measured $B_\mu/B_e$ ratio implies
then:
\be\label{eq:eta_univ}
\eta_{\tau\to l} \, = \, 0.005\pm 0.027  \ .
\ee
This determination is more accurate that the one in 
table~\ref{tab:tau_michel},
obtained from the shape of the energy distribution,
and is comparable to the value measured in $\mu$ decay.

A non-zero value of $\eta$ would show that there are at least two
different couplings with opposite chiralities for the charged leptons.
Assuming the V$-$A coupling $g_{LL}^V$ to be dominant, the
second one would be \cite{FE:90} a Higgs--type coupling $g^S_{RR}$.
To first order in new physics contributions,
$\eta\approx\mbox{\rm Re}(g^S_{RR})/2$;
Eq.~(\ref{eq:eta_univ}) puts then the (90\% CL) bound:
$-0.08 \, <\mbox{\rm Re}(g^S_{RR}) < 0.10$.

High--precision measurements of the $\tau$ decay parameters have the
potential to find signals for new phenomena. The accuracy
of the present data is still not good enough to provide
strong constraints; nevertheless, it shows
that the SM gives indeed the dominant contribution to the decay
amplitude.
Future experiments should then
look for small deviations of the SM predictions and find out the
possible source of any detected discrepancy.

In a first analysis, it seems natural to assume \cite{PS:95}
that new physics effects would be dominated by the exchange of a single
intermediate boson, coupling to two leptonic currents.
%
Table~\ref{tab:summary}
summarizes the expected changes
on the measurable shape parameters \cite{PS:95},
in different new physics scenarios.
The four general cases studied correspond to adding a single intermediate
boson exchange, $V^+$, $S^+$, $V^0$, $S^0$ 
(charged/neutral, vector/scalar), to the SM contribution
(a non-standard $W$ would be a particular case of the
SM + $V^+$ scenario).

\begin{table}[htb]
\centering
\caption{Changes in the Michel parameters induced by
the addition of a single intermediate boson exchange 
($V^+$, $S^+$, $V^0$, $S^0$)
to the SM contribution \protect\cite{PS:95}}
\label{tab:summary}
\vspace{0.2cm}
\begin{tabular}{cccccc}
\hline
&  & $V^+$  &  $S^+$  &  $V^0$  &  $S^0$
\\ \hline
$\rho - 3/4$ &     & $< 0$  &  0   &  0   &  $< 0$
\\ 
$\xi - 1$    &    &   $\pm$   & $< 0$ & $< 0$ &  $\pm$ 
\\ 
$\delta\xi-3/4$& & $< 0$ & $< 0$ & $< 0$ & $< 0$
\\ 
$\eta$        &   &   0   &  $\pm$   &   $\pm$  &  $\pm$ 
\\ \hline
\end{tabular}
\end{table}

\section*{SUMMARY}

The flavour structure of the SM is one of the main pending questions
in our understanding of weak interactions. Although we do not know the
reason of the observed family replication, we have learned experimentally
that the number of SM fermion generations is just three (and no more).
Therefore, we must study as precisely as possible the few existing flavours
to get some hints on the dynamics responsible for their observed structure.

The lepton sector provides a clean environment to test the
universality and Lorentz structure of the electroweak couplings.
We want to investigate whether the mass is the only difference
among the three fermion families.
Na\"{\i}vely, one would expect the $\tau$ to be much more sensitive
than the $e$ or the $\mu$ to new physics related to the flavour and
mass--generation problems.
While many precision measurements of the electron and muon properties
have been done in the past, it is only recently that $\tau$
experiments have achieved a comparable accuracy \cite{tau96}.
 
Lepton universality has been tested quite precisely,
both in the charged and neutral current sectors.
The leptonic couplings to the charged $W$ have been verified to be
universal at the 0.15\%  ($g_\mu/g_e$) and 0.30\% ($g_\tau/g_\mu$) level.
The axial couplings of the $Z$ boson to the charged leptons have been
measured with a comparable accuracy; universality is satisfied to
the 0.17\% ($a_\mu/a_e$) and 0.19\% ($a_\tau/a_e$) level.
The experimental precision is worse for the $Z$ vector couplings,
which are known to be the same for the three charged leptons
to 9\% ($v_\mu/v_e$) and 5\% ($v_\tau/v_e$) accuracy.

The Lorentz structure of the $l^-\to\nu_l l'^-\bar\nu_{l'}$ decay amplitudes
has been investigated by many experiments.
The present data nicely show that the bulk of the
$\mu$--decay transition amplitude in indeed of the predicted
V$-$A type.
The available information on the leptonic $\tau$ decays, is still not
good enough to determine the underlying dynamics;
nevertheless, useful limits on possible new physics contributions
start to emerge.  

At present, all experimental results  are consistent with the SM. 
There is, however, large room for improvements. Future experiments
will probe the SM to a much deeper level of sensitivity and will explore the
frontier of its possible extensions.

\section*{ACKNOWLEDGEMENTS}
I would like to thank the organizers for 
the charming atmosphere of this meeting.
I am indebted to Manel Martinez for keeping me informed about the
most recent LEP averages, and to Wolfgang Lohmann for providing 
the PAW files to generate figures~\ref{fig:mu_couplings} and 
\ref{fig:tau_couplings}.
This work has been supported in part by CICYT (Spain) under grant 
No. AEN-96-1718.



\section*{REFERENCES}
\begin{thebibliography}{99}


\bibitem{jaca:94} A. Pich, {\it The Standard Model of Electroweak
  Interactions}, Proc. XXII International Winter Meeting on Fundamental
  Physics: {\it The Standard Model and Beyond} (Jaca, 7--11 February 1994),
  eds. J.A.~Villar and A.~Morales (Editions Fronti\`eres, Gif-sur-Yvette, 
  1995), p.~1 [hep-ph/9412274].

\bibitem{sorrento:94} A. Pich, {\it Quantum Chromodynamics}, Proc. 1994
  European School of High Energy Physics 
  (Sorrento, 29 August -- 11 September 1994), eds. N.~Ellis
  and M.B.~Gavela, Report CERN 95-04 (Geneva, 1995), p.~157 
  [hep-ph/9505231].

\bibitem{comillas:95} A. Pich, {\it Flavourdynamics}, Proc. XXIII
  International Winter Meeting on Fundamental Physics: {\it The Top
  Quark, Heavy Flavour Physics and Symmetry Breaking}
  (Comillas, 22--26 May 1995), eds. T.~Rodrigo and A.~Ruiz
  (World Scientific, Singapore, 1996), p.~1 [hep-ph/9601202].

\refjl{PDG:96}{Particle Data Group}{{\it Review of Particle 
   Properties}, \PR}{D54}{1996}{1}

\bibitem{tau96} A. Pich, {\it Tau Lepton Physics: Theory Overview},
   Proc. Fourth Workshop on Tau Lepton Physics --TAU96--
   (Colorado, 16--19 September 1996), ed. J. Smith, {\it\NPPS} in press
   [IFIC/96-96; hep-ph/9612308].


\refjl{MA:94}{W.J. Marciano}{\NPPS}{40}{1995}{3} 

\bibitem{MA:90} H.-U. Martyn, {\it Test of QED by High Energy
  Electron--Positron Collisions}, in Ref.~\citenum{KI:90}.

\bibitem{KI:90} T. Kinoshita (editor), {\em Quantum Electrodynamics},
Advanced Series on Directions in High Energy Physics, Vol. 7
(World Scientific, Singapore, 1990).

\refjl{HKS:96}{M. Hayakawa, T. Kinoshita and A.I. Sanda}{\PR}{D54}{1996}{3137}
\refjl{KS:95}{T. Kinoshita and A.I. Sanda}{\PRL}{75}{1995}{790}

\refjl{BPP:96}{J. Bijnens, E. Pallante and J. Prades}{\NP}{B474}{1996}{379}

\bibitem{KR:96} B. Krause, {\it Higher--Order Hadronic Contributions to the
  Anomalous Magnetic Moment of Leptons}, Karlsruhe preprint TTP96-26 (1996).

\bibitem{AL:96} R. Alemany, {\it Tau Vector Spectral Functions and the Hadronic
  Contributions to $(g-2)_\mu$ and $\alpha(M_Z^2)$},
  Proc. Fourth Workshop on Tau Lepton Physics --TAU96--
   (Colorado, 16--19 September 1996), ed. J. Smith, {\it\NPPS} in press.

\refjl{SI:83}{\ D.J. Silverman and G.L. Shaw}{\PR}{D27}{1983}{1196}

\refjl{MA:89}{\ R. Marshall}{\RPP}{52}{1989}{1329}

\refjl{EM:93}{\ R. Escribano and E. Masso}{\PL}{B301}{1993}{419;
   {\it\NP} {B429} (1994) 19; hep-ph/9609423}

\refjl{SC:48}{\ J.S. Schwinger}{\PR}{73}{1948}{416}

\refjl{NA:78}{\ S. Narison}{\JPG}{4}{1978}{1849}
 
\refjl{SLM:91}{\ M.A. Samuel, G. Li and R. Mendel}{\PRL}{67}{1991}{668}



\refjl{KS:59}{T. Kinoshita and A. Sirlin}{\PR}{113}{1959}{1652}

\refjl{MS:88}{W.J. Marciano and A. Sirlin}{\PRL}{61}{1988}{1815}

\bibitem{PI:92} A. Pich, {\it Tau Physics}, in {\it Heavy Flavours}, 
  eds. A.J.~Buras and M.~Lindner,
  Advanced Series on Directions in High Energy Physics -- Vol.~10
  (World Scientific, Singapore, 1992), p.~375.

\refjl{MS:93}{W.J. Marciano and A. Sirlin}{\PRL}{71}{1993}{3629}

\refjl{DF:94}{R. Decker and M. Finkemeier}{\NP}{B438}{1995}{17;
         {\it\NPPS} {40} (1995) 453}

\refjl{BR:92}{D.I. Britton \etal}{\PRL}{68}{1992}{3000}

\refjl{CZ:93}{G. Czapek \etal}{\PRL}{70}{1993}{17}

\refjl{UA1:89}{C. Albajar \etal\ (UA1)}{\ZP}{C44}{1989}{15}
\refjl{UA2:92}{J. Alitti \etal\ (UA2)}{\PL}{B280}{1992}{137}
\refjl{CDF:92}{F. Abe \etal\ (CDF)}{\PRL}{68}{1992}{3398;
      {69} (1992) 28}
 

\bibitem{LEP:96}
 The LEP Electroweak Working Group and the SLD Heavy Flavour Group,
{\it A Combination of Preliminary LEP and SLD
Electroweak Measurements and Constraints on the Standard Model},
CERN preprint LEPEWWG/96-02 (30 July 1996).

\refjl{BPR:91}{J. Bernab\'eu, A. Pich and N. Rius}{\PL}{B257}{1991}{219} 

\bibitem{FSanchez} F. S\'anchez, {\it Measurement of the Transverse
  Spin Correlations in the Decay $Z\to\tau^+\tau^-$}, 
   Proc. Fourth Workshop on Tau Lepton Physics --TAU96--
   (Colorado, 16--19 September 1996), ed. J. Smith, {\it\NPPS} in press.
   

\refjl{ABGPR:92}{R. Alemany \etal}{\NP}{B379}{1992}{3}

\refjl{ALEPH:94}{D. Buskulic \etal\ (ALEPH)}{\PL}{B321}{1994}{168}

\refjl{L3:96}{M. Acciarri \etal\ (L3)}{\PL}{B377}{1996}{313} 

\refjl{CHARMII:94}{P. Vilain \etal\ (CHARM II)}{\PL}{B335}{1994}{246}

\refjl{CHARMII:94b}{P. Vilain \etal\ (CHARM II)}{\PL}{B320}{1994}{203}


\refjl{MI:50}{L. Michel}{Proc. Phys. Soc.}{A63}{1950}{514; 1371}
 
\refjl{BM:57}{C. Bouchiat and L. Michel}{\PR}{106}{1957}{170; \\
   T. Kinoshita and A. Sirlin, {\it\PR} 107 (1957) 593; 108 (1957) 844}


\bibitem{SCH:83} F. Scheck, {\it Leptons, Hadrons and Nuclei}
   (North-Holland, Amsterdam, 1983);
   {\it\PRep} {44} (1978) 187.

\refjl{FGJ:86}{W. Fetscher, H.-J. Gerber and K.F. Johnson}{\PL}{B173}
    {1986}{102}

\bibitem{FG:93} W. Fetscher and H.-J. Gerber, {\it Precision
  Measurements in Muon and Tau Decays}, in {\it Precision Tests of the
  Standard Electroweak Model}, ed. P.~Langacker,
  Advanced Series on Directions in High Energy Physics -- Vol.~14
  (World Scientific, Singapore, 1995), p.~657.
  
\refjl{PS:95}{A. Pich and J.P. Silva}{\PR}{D52}{1995}{4006}

\bibitem{Rouge} A. Roug\'e, {\it Results on $\tau$ Neutrino from Colliders},
 Proc. XXX$^{\mbox{\rms th}}$ Rencontre de Moriond, 
 {\it Dark Matter in Cosmology, Clocks and Tests of Fundamental Laws}, 
eds. B. Guiderdoni \etal\ (Editions Fronti\`eres, Gif-sur-Yvette, 1995)
p.~247.

\refjl{JA:66}{C. Jarlskog}{\NP}{75}{1966}{659}

\refjl{FE:84}{W. Fetscher}{\PL}{140B}{1984}{117}
%

\refjl{TS:71}{Y.S. Tsai}{\PR}{D4}{1971}{2821; \\
%
   S. Kawasaki, T. Shirafuji and S.Y. Tsai, {\it Progr. Theor. Phys.}
   {49} (1973) 1656}

\refjl{PS:77}{S.-Y. Pi and A.I. Sanda}{\APNY}{106}{1977}{171}


\bibitem{NE:91} C.A. Nelson, {\it\PR} {D43} (1991) 1465; 
   {\it\PRL} {62} (1989) 1347; {\it\PR} {D40} (1989) 123 
    [{\it Err:} {D41} (1990) 2327].
%
\bibitem{GN:91} 
   S. Goozovat and C.A. Nelson, {\it\PR} {D44} (1991) 2818; 
   {\it\PL} {B267} (1991) 128 [{\it Err:} {B271} (1991) 468].

\refjl{FE:90}{W. Fetscher}{\PR}{D42}{1990}{1544}

\refjl{DDDR:93}{M. Davier \etal}{\PL}{B306}{1993}{411}

\bibitem{SV:96} A. Stahl and H. Voss, BONN-HE-96-02.

\bibitem{LR:95} W. Lohmann and J. Raab, {\it Charged Current Couplings in
  $\tau$ Decay}, DESY 95-188.

\refjl{ST:94}{A. Stahl}{\PL}{B324}{1994}{121}

 




\end{thebibliography}
\end{document}